\documentclass[11pt]{scrartcl}
\newcommand{\topicnumber}{A3}
\newcommand{\icpigheader}{30$^\text{th}$ ICPIG, August 28$^\text{th}$ - Septemper 2$^\text{nd}$, 2011, Belfast, UK}
\usepackage{cite}
\usepackage{amsmath,amssymb}
\usepackage{ulem}
\usepackage{graphicx,xcolor}
\xdefinecolor{headercolor}{RGB}{128,128,128}
\usepackage[small]{caption2}
\usepackage[paper=a4paper,left=20mm,right=20mm,top=25mm,bottom=25mm]{geometry}
\newfont{\titelfont}{cmbx10 scaled 1400}
\newfont{\authorfont}{cmr10 scaled 1200}
\newfont{\institutefont}{cmsl10 scaled 1000}
\newfont{\abstractfont}{cmr10 scaled 1000}
\newfont{\captionfontICPIG}{cmr10 scaled 1000}
\newfont{\headerfont}{cmr10 scaled 800}

\newlength\abstractwidth
\newcommand{\presenter}[1]{\uline{#1}}

\makeatletter
	\renewcommand*\bib@heading{}
\makeatother
\usepackage{fancyhdr}
\providecommand{\eqref}[1]{(\ref{#1})}

\newcommand{\abs}[1]{\left\vert#1\right\vert}

\newcommand{\eps}{\varepsilon}

\newcommand{\phdag}{{\phantom{\dag}}}

\newcommand{\conj}{{*}}
\newcommand{\phconj}{{\phantom{\conj}}}

\newcommand{\jprime}{{\prime}}

\newcommand{\jvecprime}{{\,\jprime}}

\newcommand{\NitrogenGroundStateTermSymbol}{\ensuremath{^1\Sigma_g^+}}
\newcommand{\NitrogenGroundState}{\ensuremath{N_2(\NitrogenGroundStateTermSymbol)}}

\newcommand{\NitrogenDominantMetastableStateTermSymbol}{\ensuremath{^3\Sigma_u^+}}
\newcommand{\NitrogenDominantMetastableState}{\ensuremath{N_2(\NitrogenDominantMetastableStateTermSymbol)}}

\newcommand{\NitrogenNegativeIonResonanceTermSymbol}{\ensuremath{^2\Pi_g}}
\newcommand{\NitrogenNegativeIonResonance}{\ensuremath{N_2^-(\NitrogenNegativeIonResonanceTermSymbol)}}

\newcommand{\PenningME}[1]{V_{\scriptscriptstyle 0#1, \vec{k}}^{\scriptscriptstyle \vec{q}, 1#1}}

\newcommand{\citeAuthorStracke}{Stracke et al.}
\newcommand{\citeAuthorMakoshi}{Makoshi}

\begin{document}
\sloppy
\twocolumn[
\center{
{\titelfont Secondary electron emission due to Auger de-excitation of metastable nitrogen molecules at metallic surfaces}\\[0.5cm]

{\authorfont J. Marbach, \presenter{F. X. Bronold}, H. Fehske}\\[0.5cm]

{\institutefont Institut f\"ur Physik, Ernst-Moritz-Arndt-Universit\"at Greifswald, 17489 Greifswald, Germany}
}

\center{\parbox[b]{\abstractwidth}{\abstractfont
With an eye on plasma walls we investigate, within an effective model for the 
two active electrons involved in the process, secondary electron emission due 
to Auger de-excitation of metastable nitrogen \NitrogenDominantMetastableState\ 
molecules at metallic surfaces. Modelling bound and unbound molecular 
states by a LCAO approach and a two-center Coulomb wave, respectively, and the metallic 
states by the eigenfunctions of a step potential we employ Keldysh Green's 
functions to calculate the secondary electron emission coefficient $\gamma_e$ retaining for
the Auger self-energy the non-locality in time and the dependence on the single particle quantum 
numbers. For the particular case of a tungsten surface we find good agreement with experimental 
data indicating that the relevant Auger physics is well captured by our easy-to-use model.
}\vspace*{0.5cm}}]

\noindent{\bf 1. Motivation}\\\indent
Secondary electron emission by surface scattering of metastable molecules is an important process in molecular low-temperature gas discharges. It is one of the main wall-based electron generation channels controlling, together with wall recombination and various volume-based charge production and destruction channels, the overall charge balance in the discharge. The efficiency of electron release can be encapsulated in the secondary electron emission coefficient $\gamma_e$, which denotes the probability for releasing an electron by a single metastable de-excitation at the wall. However, for most wall materials and projectiles $\gamma_e$ is unknown. A flexible, easy-to-use microscopic approach for its calculation is thus needed.

\citeAuthorStracke\cite{stracke98} experimentally investigated the de-excitation of metastable nitrogen 
\NitrogenDominantMetastableState\ molecules on a tungsten surface and proposed two reaction channels. 
Firstly, Auger de-excitation (also referred to as Penning de-excitation when the emitted electron 
comes from the molecule), $\NitrogenDominantMetastableState + e_m \rightarrow \NitrogenGroundState + e_f$, and, 
secondly, the formation of a \NitrogenNegativeIonResonance\ shape resonance with subsequent auto-detachment, 
$\NitrogenDominantMetastableState + e_m \rightarrow \NitrogenNegativeIonResonance \rightarrow \NitrogenGroundState + e_f$, 
where $e_m$ and $e_f$ denote an electron inside the metal and a free electron, respectively. 
\citeAuthorStracke\cite{stracke98} concluded that the latter process should be more efficient, as it is a combination 
of two single-electron charge-transfer transitions, whereas the Auger reaction represents a less probable two-electron 
transition. Using thermal molecules they measured the energy spectrum of the released electron and estimated 
$\gamma^{\rm exp}_e|_{\rm total}$ to be about $10^{-3}-10^{-2}$. This value does however not discriminate between 
the two channels.
Indeed, in the energy-resolved data, \citeAuthorStracke\cite{stracke98} also find a weak signal due to Auger
de-excitation giving rise to $\gamma^{\rm exp}_e|_{\rm Auger}\approx 10^{-4}-10^{-3}$. In view of the 
electronic structure of tungsten this is not surprising because large energy shifts 
and broadening due to image interactions are required to bring the molecular orbital hosting the 
hole in the electronic configuration of \NitrogenDominantMetastableState\ in resonance with conduction band 
states. Charge transfer may thus be less efficient and Auger de-excitation an eye-to-eye 
competitor.

To establish our approach and because of the availability of experimental data 
we focus in the present work on Auger de-excitation of \NitrogenDominantMetastableState\ 
at an uncharged tungsten surface with simultaneous release of an electron. We construct an effective model that 
concentrates on the
most important degrees of freedom and enables us to describe the system by a few parameters which are 
accessible through experiments or theoretical calculations. The primary goal will be to calculate 
$\gamma_e$ using Keldysh Green's functions along the lines layed out by 
\citeAuthorMakoshi~\cite{makoshi91} in his investigation of de-excitation of metastable atoms. 
\\

\begin{figure}[t]
        \center
        \includegraphics[clip,width=0.98\linewidth]{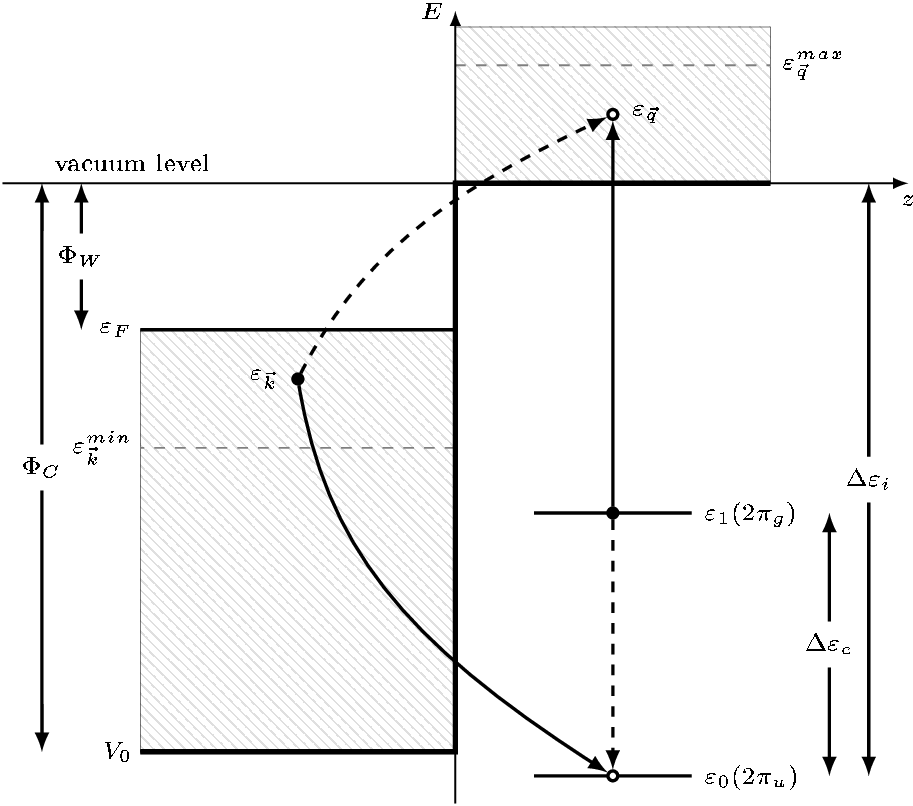}
        \caption{Illustration of the effective model showing Penning de-excitation (solid lines) and its exchange process (dashed lines). Here $\eps_{\vec{k}}^{min}$ and $\eps_{\vec{q}}^{max}$ are the classical energy cut-offs obtained from the energy balance $\eps_1 + \eps_{\vec{k}} = \eps_0 + \eps_{\vec{q}}$ that holds in the adiabatic limit.}
        \label{energy scheme}
\end{figure}

\noindent{\bf 2. Model}\\\indent
Focussing on the essentials of the process, we assume the metallic surface to be planar, ideal, 
and to stretch over the entire half space~${z<0}$. Furthermore, we consider only the dominant metastable state \NitrogenDominantMetastableState\ and employ the trajectory approximation, that is, we decouple the translational motion of the molecule from the dynamics of the system and externally supply its trajectory. Finally, the molecule is assumed to impact the surface under normal incidence with constant velocity $v$ and constant angle $\varphi$ of its axis to the surface. Assuming the molecule to start moving at $t_0=-\infty$ and to hit the surface at $t=0$ the trajectory of its center of mass is 
$\vec{R}(t)=\left( v |t| + z_0 \right) \vec{e}_z \,,$ where $z_0$ is the turning point of the
molecule in the surface potential $V_S(z)$. We consider only indirect de-excitation (Penning process) in which the electron 
is emitted from the molecule (solid lines in Fig.~\ref{energy scheme}). Direct de-excitation
(dashed lines in Fig.~\ref{energy scheme}) can be shown to be negligible because the orthogonality of the bound 
molecular wave functions makes the corresponding matrix element much smaller than the one for Penning de-excitation.

To construct the Hamiltonian we combine three different kinds of single-electron states to a single-electron basis: the single-electron states of the conduction band of the solid surface $|\vec{k}\rangle$, which we approximate by the states corresponding to an electron trapped by a step potential of depth $\Phi_C$, the free single-electron states $|\vec{q}\rangle $ associated with the molecule's continuum for which we use a two-center Coulomb wave,~\cite{joulakian96} and effective single-electron states $|0m/1m\rangle$ (${m=\pm1}$ is the magnetic quantum number) for the bound states of the molecule, which we approximate by a degenerate two-level system keeping, within the LCAO representation of the nitrogen molecule, only the $2\pi_u$ and the $2\pi_g$ molecular orbitals (MOs) which are the two MOs whose occupancies change during the de-excitation process. The LCAO MOs are constructed by hydrogen-like wave functions with effective nucleus charges to mimic the Roothaan-Hartree-Fock wave functions of atomic nitrogen. Note, the Penning process does not involve any spin flip, we thus ignore 
the spin.

The description of the electronic structure of the molecule-surface system requires to align the single-electron states 
against each other and against the vacuum level by use of the metal's work function $\Phi_W$, the metal's conduction band 
depth $\Phi_C$, the molecule's ionization energy $\Delta\eps_i$, and the excitation energy of the molecule $\Delta\eps_e$. 
The metal states are of course occupied up to the Fermi level $\eps_F$. Our model is thus characterized by a few energy 
parameters, an effective charge, and a bond length (which enters the molecular wave functions).

Due to the symmetries of the molecular ground and excited state, only transitions with $\Delta m = 0$ are involved. 
They are driven by the Coulomb interaction between the excited electron in the $2\pi_g$ MO and an electron in the Fermi 
sea of the metal. The three electrons in the $2\pi_u$ MO act only as spectators and can thus be neglected. Assuming 
moreover the Fermi surface of the metal to be rigid, the de-excitation of \NitrogenDominantMetastableState~is a 
two-body scattering process, whose Hamiltonian, written in the single-electron basis just described, is given by 
\begin{equation}
\begin{split}
	H & = \sum_{\vec k} \eps_{\vec k}^{\phdag} \, c_{\vec{k}}^{\dag} \, c_{\vec{k}}^{\phdag} + \sum_{\vec q} \eps_{\vec q}^{\phdag} \, c_{\vec{q}}^{\dag} \, c_{\vec{q}}^{\phdag} \\
	& + \sum_{m} \eps_0^{\phdag} \, c_{0m}^{\dag} \, c_{0m}^{\phdag} + \sum_{m} \eps_1^{\phdag} \, c_{1m}^{\dag} \, c_{1m}^{\phdag} \\
	& + \sum_{\vec k, \vec q, m} \left( \PenningME{m}(t) \, c_{0m}^{\dag} c_{\vec{k}}^{\phdag} \, c_{\vec{q}}^{\dag} \, c_{1m}^{\phdag} \,+\, H.c. \right)\,.
\end{split}%
\label{Hmodel}
\end{equation}
The non-equilibrium character of the system, and hence the need to employ Keldysh Green's 
functions~\cite{makoshi91}, arises from the time-dependent Penning matrix element,
\begin{equation}
\begin{split}
	\PenningME{m}(t) = & \int \hspace{-0.3em} d\vec{r} \hspace{-0.1em} \int \hspace{-0.3em} d\vec{r}^\jvecprime \; \Psi_{0m}^{\conj} \bigl( \vec{r}_\varphi(t) \hspace{-0.1em} \bigr) \Psi_{\vec{k}}^{\phconj} \bigl( \vec{r}\, \bigr) \\
	& \times V_C \bigl( \abs{\vec{r} - \vec{r}^\jvecprime} \bigr) \Psi_{\vec{q}_\varphi}^{\conj} \bigl( \vec{r}_\varphi^\jvecprime(t) \hspace{-0.1em} \bigr) \Psi_{1m}^{\phconj} \bigl( \vec{r}_\varphi^\jvecprime(t) \hspace{-0.1em} \bigr)~,
\end{split} \label{penning matrix element}
\end{equation}
where $V_C$ represents the Coulomb potential and the subscript $\varphi$ denotes the associated vector as seen from 
the molecule's reference frame. 

Image interactions are not considered except for the emitted electron, which always feels the image potential 
$V_I(z)$ and can only escape from the surface when its perpendicular energy is sufficiently high. To take this
into account we multiplied in the calculation of $\gamma_e$ and the spectrum of the emitted electron
the Penning matrix element by the surface transmission function $\Theta(\eps_{q_z} - V_I(z_R(t)))$.\\

\noindent{\bf 3. Quantum kinetics}\\\indent
The secondary electron emission coefficient can be obtained from the time-dependence of the
occupancy of the free states $|\vec{q}\rangle$. To calculate this quantity from~(\ref{Hmodel}) 
we employ Keldysh Green's functions along the lines pioneered by \citeAuthorMakoshi.~\cite{makoshi91} 
In contrast to him, however, we do not work with a phenomenological Auger interaction 
and do not employ the wide-band approximation for the free states. 
We are also not restricted to the lowest order expressions for 
the occupancies (given below by \eqref{n1 result lowest order}, \eqref{nq result}, and 
\eqref{gamma e 0 result}), but can in principle calculate higher order corrections to these 
expressions. 

The central quantity of our approach (in detail described in Ref.~\cite{marbach11}) is 
a function $\Delta_m(t_1,t_2)$
\begin{equation}
\begin{split}
	& \Delta_m(t_1,t_2) = \frac{1}{\hbar^2} \sum_{\vec{q},\vec{k}} \left[ \PenningME{m}(t_1) \right]^\conj \PenningME{m}(t_2) \\
	& \qquad \times n_{\vec{k}}(t_0) \, e^{-\frac{i}{\hbar} \left( \eps_0 + \eps_q - \eps_1 - \eps_k \right) (t_1 - t_2)}~,
\end{split}\label{delta 1m definition}
\end{equation}
which emerges from the self-energy terms of the iterated Dyson equation. To lowest order in $\Delta_m$ the occupation of the 
excited molecular level is
\begin{equation}
	n_{1\mu}(t) = e^{- \hspace{-0.15em} \int_{t_0}^t \hspace{-0.3em} dt_1 \hspace{-0.1em} \int_{t_0}^t \hspace{-0.3em} dt_2 \, \Delta_\mu(t_1,t_2)}~,
	\label{n1 result lowest order}
\end{equation}
with $\mu$ being the initial magnetic quantum number of the excited electron. For the spectrum of emitted electrons we obtain 
to lowest order in $\Delta_m$
\begin{equation}
	n_{\vec{q}}(t) = \int_{t_0}^{t^{\phantom{\jprime}}} \hspace{-0.6em} dt_1 \hspace{-0.2em} \int_{t_0}^{t^{\phantom{\jprime}}} \hspace{-0.6em} dt_2 \; {\Delta}_{\mu,\vec{q}}(t_1,t_2)~,
	\label{nq result}
\end{equation}
where ${\Delta}_{\mu,\vec{q}}$ is defined by Eq.~\eqref{delta 1m definition} without the sum over $\vec{q}$.

The secondary electron emission coefficient $\gamma_e$ can be calculated from Eq.~\eqref{nq result} by taking $t=\infty$ and summing over all possible $\vec{q}$, which to lowest order in $\Delta_m$ yields
\begin{equation}
	\gamma_e^{(0)} = \int_{t_0}^{\infty} \hspace{-0.6em} dt_1 \hspace{-0.2em} \int_{t_0}^{\infty} \hspace{-0.6em} dt_2 \; {\Delta}_{\mu}(t_1,t_2)~,
	\label{gamma e 0 result}
\end{equation}
where $\mu$ is again the initial magnetic quantum number.
If we allow for the next order terms with respect to $\Delta_m$ in \eqref{nq result} we find
\begin{equation}
	\gamma_e^{(1)} = 1 - e^{- \hspace{-0.15em} \int_{t_0}^\infty \hspace{-0.3em} dt_1 \hspace{-0.1em} \int_{t_0}^\infty \hspace{-0.3em} dt_2 \, \Delta_\mu(t_1,t_2)}~.
    \label{gamma e 1 result}
\end{equation}
However, the zeroth and first order results \eqref{gamma e 0 result} and \eqref{gamma e 1 result} differ significantly 
only for very small kinetic energies of the molecule.~\cite{marbach11}

To simplify the numerical analysis of \eqref{n1 result lowest order}--\eqref{gamma e 1 result} we utilize the localized nature of the molecular wave functions, which together with the fact, that the molecule's turning point $z_0$ lies far outside the surface ($z_0 \geq 4.35\,a_B$) allows us to neglect the oscillating part of the metal wave functions inside the solid and to separate the time dependence of the Penning matrix element.~\cite{marbach11} The time integrals of $\Delta_m(t_1,t_2)$ can then be calculated analytically. The remaining wave vector dependent part of the matrix element is calculated on a discrete grid within the {$(\vec{k},\vec{q}\hspace{0.1em})$-space} and afterwards the wave vector integrals within $\Delta_m$ are calculated using sexa-linear interpolation to obtain the inter-grid values of $\PenningME{m}$.\\

\noindent{\bf 4. Results}\\\indent
We now present selected results for the particular case of a tungsten surface. The kinetic energy of the molecule was fixed to $50\,meV$. The associated turning point $z_0$ amounts to approximately $4.42\,a_B$. We did not find any influence of the initial magnetic quantum number $\mu=\pm 1$. For convenience we thus omit any $m$ subscripts in the following. In addition we restrict the molecule's orientation to the two fundamentally distinct situations $\varphi=0$ (axis parallel to the surface) and $\varphi=\frac{\pi}{2}$ (axis perpendicular to the surface).

\begin{figure}[t]
\center
\includegraphics[clip,width=0.98\linewidth]{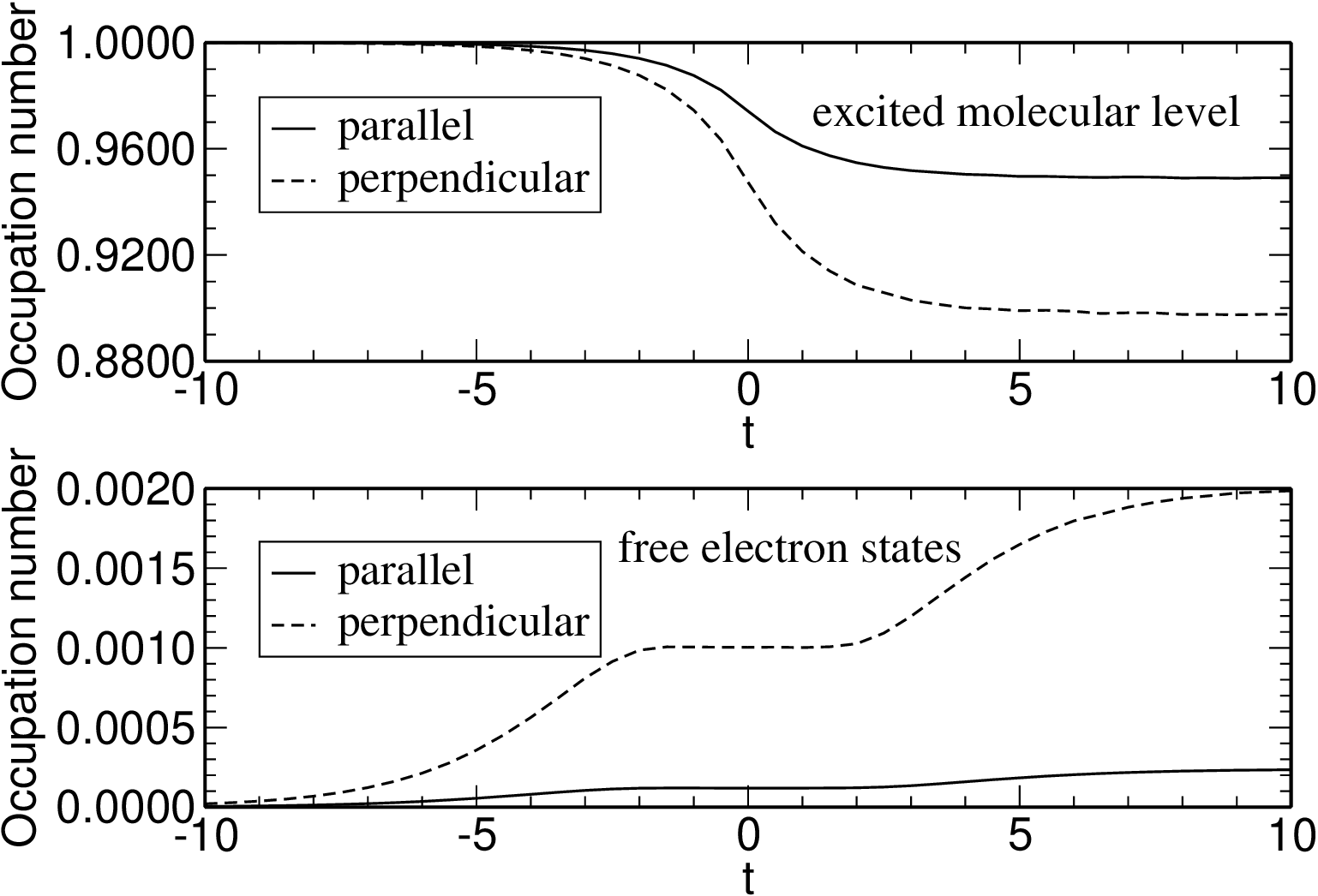}
\caption{Time evolution of the occupancies of the excited molecular level and the free electron states for parallel (solid lines) and perpendicular (dashed lines) molecule orientation.}
\label{dynamics fig}
\end{figure}

Figure~\ref{dynamics fig} shows the time evolution of the occupancies of the excited molecular level and the free electron states. Obviously, the occupancy of the excited molecular level (upper panel) changes significantly only in the range $\abs{t}\leq 5$, which, taking the turning point into account, equals maximum distances of the molecule's center of mass from the surface of roughly $7\,a_B$. 
Within our model, the process is equally effective in the incoming and outgoing branch of the trajectory. The time evolution of the occupancy of the free states (lower panel) is however distinctively different. It shows a plateau around $t=0$, that is, a stagnation of the probability to escape from the solid. This is a consequence of the image potential which almost completely traps an electron emitted at low surface distances where its perpendicular energy is too small to overcome the image barrier encoded in the surface transmission function. The number of emitted electrons is larger in the perpendicular case as compared to the parallel case. Numerical tests showed that for the parallel case the electron is primarily emitted with a very small perpendicular energy whereas in the perpendicular case the perpendicular energy of the emitted electron is distributed more equally. Thus, in the perpendicular orientation the electron has a higher probability to breach through the surface barrier originating from the image potential. 
The occupancy of the free states for $t\rightarrow\infty$, that is, the secondary electron emission coefficient 
amounts to $2\cdot 10^{-3}$ in the perpendicular orientation and $2.4 \cdot 10^{-4}$ in the parallel orientation which is quite 
close to the experimental estimate given
by Stracke and coauthors, $\gamma^{\rm exp}_e|_{\rm Auger}\approx 10^{-4}-10^{-3}$.~\cite{stracke98}

\begin{figure}[t]
\center
\includegraphics[clip,width=0.98\linewidth]{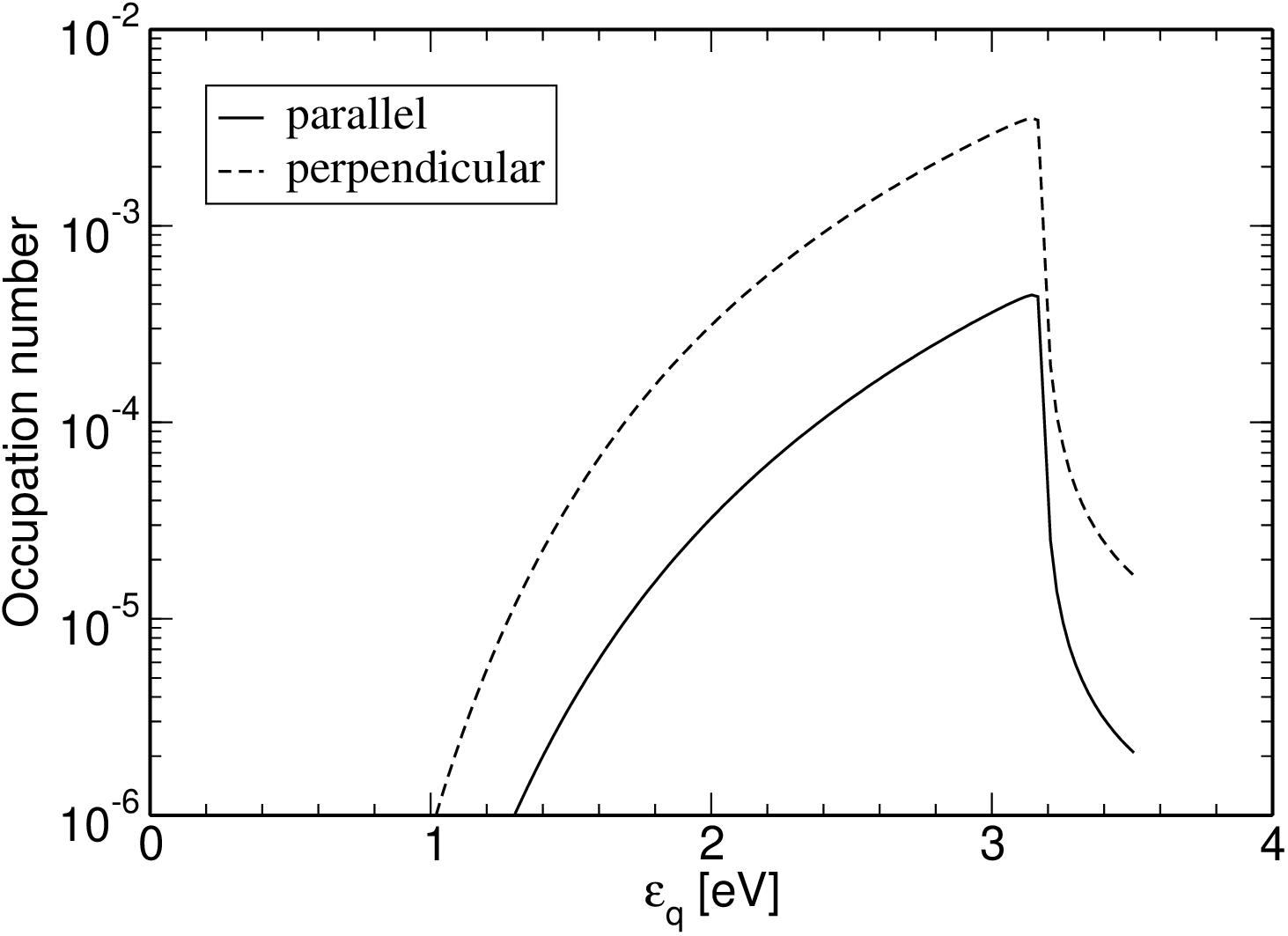}
\caption{Spectrum of the emitted electron for parallel (solid lines) and perpendicular (dashed lines) molecule orientation.}
\label{spectrum fig}
\end{figure}

Figure~\ref{spectrum fig} finally shows the spectrum of the emitted electron at $t=\infty$. The graphs for the two different 
orientations start at the origin and monotonously increase until a cut-off energy is reached. The latter resembles 
the classical cut-off energy $\eps_{\vec{q}}^{max}$ (see Fig.~\ref{energy scheme}), implying that energy conservation is 
restored at the end of the collision, as it should be. The low energy part of the spectrum is cut off due to the surface 
transmission function which allows electrons to escape from the surface only when their perpendicular energy is large enough. 
The spectrum for the perpendicular case takes on larger values, for reasons already explained.\\

\noindent{\bf 5. Conclusions}\\\indent
We investigated within the Keldysh formalism the release of secondary electrons due to Auger de-excitation 
of \NitrogenDominantMetastableState\ at metallic surfaces using an effective model for the two active 
electrons involved in the process. For tungsten we obtained 
for this particular electron emission channel $\gamma_e\approx 2\cdot10^{-3}$ 
(perpendicular orientation) and $\gamma_e\approx 2.4 \cdot10^{-4}$ (parallel orientation) which agrees well with 
experimental estimates. The effective model depends only on a few, easily obtainable energies -- 
a real advantage for applications to low-temperature gas discharges, where a great variety of different
kinds of molecules and wall materials occur. With some modifications the model can be applied to dielectric 
surfaces as well. We employed the model only for the investigation of Auger de-excitation. But provided the 
molecule's image interactions and the on-molecule Coulomb interaction between the two active electrons are 
taken into account, it can be also used to study charge-transfer processes.\\

\end{document}